# Malicious Code Detection: Run Trace Output Analysis by LSTM

CENGIZ ACARTURK, (Member, IEEE), MELIH SIRLANCI, PINAR GURKAN BALIKCIOGLU,
DENIZ DEMIRCI, NAZENIN SAHIN, AND OZGE ACAR KUCUK
Informatics Institute, Middle East Technical University, 06800 Ankara, Turkey

Corresponding author: Cengiz Acarturk (acarturk@metu.edu.tr)

This work was partially supported by Middle East Technical University funds.

**ABSTRACT** Malicious software threats and their detection have been gaining importance as a subdomain of information security due to the expansion of ICT applications in daily settings. A major challenge in designing and developing anti-malware systems is the coverage of the detection, particularly the development of dynamic analysis methods that can detect polymorphic and metamorphic malware efficiently. In the present study, we propose a methodological framework for detecting malicious code by analyzing run trace outputs by Long Short-Term Memory (LSTM). We developed models of run traces of malicious and benign Portable Executable (PE) files. We created our dataset from run trace outputs obtained from dynamic analysis of PE files. The obtained dataset was in the instruction format as a sequence and was called Instruction as a Sequence Model (ISM). By splitting the first dataset into basic blocks, we obtained the second one called Basic Block as a Sequence Model (BSM). The experiments showed that the ISM achieved an accuracy of 87.51% and a false positive rate of 18.34%, while BSM achieved an accuracy of 99.26% and a false positive rate of 2.62%.

## I. INTRODUCTION

Today's evolving information systems are frequently attacked with malicious intent or different motivations. Since the development of the systems makes the attack surface bigger, the number of attacks increases each day. One of the main attack methods is malicious software, i.e., malware, which includes specific types such as viruses, worms, and trojans. Malware can be used to attack operating systems and applications and cause damage at both personal and corporate levels. Usually, exploiting a vulnerability in computer systems through malicious software, real-time systems' availability is targeted, and valuable data is rendered unusable. The spread of this type of malware is becoming faster due to the increased connectivity of new devices such as computers, smartphones, smart televisions, and devices in the home area network, i.e., IoT devices. Besides, the increase in mobile devices' use encourages malware authors to focus on mobile operating systems and applications, which will eventually lead to an expansion of malware detection and mitigation methodologies into novel domains of application.

Over the past decade, the number of new malware obtained daily has been increasing. According to the IT Security Institute, AV-TEST statistics, 350,000 new malware and unwanted applications are examined and classified every day (August 2020) [1]. Besides, the online malware analysis service Virustotal reports statistical data on files submitted for analysis [2]. According to these statistics, the average daily number of files sent for analysis was 2 million in the seven days between July 28 and August 4, 2020. The average number of unique files submitted was 1.6 million, of which 800,000 were detected daily by one or more AV(Antivirus) engines. Another vital point drawn from the statistics for the 7 days is that 4.4 million of the files sent during the week were x86 Windows Operating system executable files. The daily amount of suspicious submitted files is continuously increasing. This situation brings the need for a richer set of methodologies for malware analysis. In the present study, we aim to enrich the malware analysis methodology by proposing a framework for an automated run trace output analysis, a recent challenge in cybersecurity defense systems.

  

Traditional methods can no longer perform well on polymorphic and metamorphic types of malware recently. Polymorphic malware uses encryption to escape from AV products. Usually, it has decryption modules built-in, so it is on disk in encrypted form. This software only decrypts and executes malicious parts of it at runtime using decryption modules. When such malware is on the disk, it has a benign appearance as perceived by the host computer and, therefore, can bypass static AV scanners without changing its appearance.

On the other hand, metamorphic malware can change its look further on each execution. It uses obfuscation techniques to change the appearance and obtain functionally equivalent versions of itself. Since a metamorphic engine produces a malware file that does the same job in each execution but looks different from the previous one, it becomes virtually impossible to detect by signature-based methods. It also causes an increase in the number of malware discovered daily, as new copies of malware are obtained using polymorphic and metamorphic types of malware. Because such malware is difficult to detect manually, the research focus has shifted to the use of Machine Learning (ML) in automated malware detection systems.

Malware analysis is usually divided into two main categories, namely static analysis and dynamic analysis. The static analysis aims at gathering information about a suspected file without executing it to decide whether it is malicious or not. During static analysis, analysts often use a disassembler tool and investigate the assembly code, imported functions, and strings. On the other hand, in dynamic analysis, the suspected file is executed, and information about likely malicious operations is collected. Thus, a dynamic analysis should be performed by isolating the run file in a controlled environment such as sandboxes or virtual machines to avoid possible infections. During dynamic analysis, the program's flow is traced, and the malware analysts examine the function calls and the parameter values in registers. As in malware analysis, malware detection research by using machine learning and deep learning focuses on similar data collected from files such as assembly code, opcodes, API (Application Programming Interface) calls, control flow graphs, and metadata from file headers, e.g., [3]–[5].

Machine learning and deep learning techniques are used to detect malware in various fronts, such as conducting binary classification of software as benign or malicious, as well as classifying malware into known types such as virus, worm, and trojan or known malware families. In our study, we focused on deep learning methods and then used a specialized type of Recurrent Neural Network (RNN) called Long Short-Term Memory (LSTM) proposed by Hochreiter and Schmidhuber [6]. We approach malware detection from the perspective of Natural Language Processing (NLP) by developing and testing models that process run traces of malicious and benign software.

We propose a novel approach to malware detection that focuses on run trace components in a dynamic analysis framework. Recently, there exists a limited number of studies using dynamic analysis with assembly instructions. No study uses the run trace output for malware detection to the best of our knowledge. In the present study, we aim at exploring the detection performance of dynamically collected data. We report an investigation of run trace data collected at runtime of PE (Portable Executable) files. In particular, we used a semi-automated process to collect run trace output from PE files. First, we created the run trace dataset as an instruction per line, viz. instruction as a sequence. Then, we converted it into a different form as a basic block per line, viz. basic block as a sequence; thus, we obtained a second dataset. After creating our datasets, we chose LSTM as the machine learning technique. As reported in the literature [7] and [8], LSTM shows better performances than its predecessor, RNN (also among customized versions of RNN). We called our proposed methods "ISM (Instruction as a Sequence Model)" and "BSM (Basic Block as a Sequence Model)." We aim to compare ISM and BSM based on the evaluation results. The following sections present the methodological details.

We decided to focus on the malware detection problem since it is an increasingly severe threat to information systems. Language modeling and text classification approaches can be useful to solve this problem, so we adapted them into malware detection context. We used Long Short-Term Memory (LSTM) to do binary classification on dynamically collected assembly instructions.

Deep Learning methods provide more resistance against changes in data since feature extraction is automatically handled by neural networks instead of manual feature extraction in Machine Learning. Besides, modeling and classification of natural languages by employing various types of RNN were shown to achieve high accuracies in previous works [7], [8]. Among standard RNN and its specialized versions, we preferred to use LSTM since it provides more robust architecture during the training phase by better solving vanishing and exploding gradient problems of standard RNN architecture. Because of the structural and semantic similarities between a natural language and assembly language, to detect malicious software, we decided to create language models of assembly instructions from malicious and benign executable files as two different natural languages. Also, we worked on two different forms of the same assembly instruction data to find out the effect of structural and semantic differences of assembly code on the detection capability and whether handling data differently achieves better results.

The paper is organized as follows. In the next section, we present the studies about malware detection. Next, we describe our approach, datasets, parameters, and the proposed models (viz. ISM and BSM). Then, we report the results and a comparison of the models. Finally, we present a discussion of the results and the limitations of the study.

## II. RELEVANT WORK
The major challenge in today's cybersecurity strategies is that malware developers continuously update their methodologies, thus generating novel malware types, which are difficult to detect by the automated analysis tools. In particular, integrating artificial intelligence and machine learning into mitigation techniques aims to develop malware detection systems with high accuracy, low false-positive rates, and best performance. A review of the literature reveals the use of three major features for the development of automatic malware detection systems: opcode, frequency, and sequence,[1] Application Programming Interface (API) calls, and Control Flow Graphs (CFG). Our study examines assembly instructions obtained from run trace outputs of PE (Portable Executable) files.

Opcode frequency and sequence, usually obtained from the static analysis processes, comprise the backbone of any program code syntax. Therefore, they can be used as features for malware detection. The opcode sequences provide valuable information about *semantic* aspects of the program codes (as described within the framework of word embedding models employed for Natural Language Processing).

Since the past decade, the study of opcodes for malware analysis has been subject to various methodological analyses. For example, Bilar [9] performed the extraction of common and rare opcodes from PE files using descriptive statistics, specifically to classify certain types of malware such as trojans and worms. Santos *et al.* [10] studied the incidence of opcode sequences. They investigated the relationships among the opcodes and used statistical information to detect variants of known malware families.

Machine learning for malware detection and classification has been popular for the past decade on various fronts. For example, in [11], opcode frequency has been used as the main feature of the ML model, obtained from the assembly output of executables. The ML methodologies included Support Vector Machine (SVM), Random Forest (RF), Decision Tree (DT), and BOOSTING, among others, for classifying executables as malicious or benign. In addition to ML modeling by independent opcode features, Shabtai *et al.*'s investigation of n-gram opcode sequences has enriched automatic detection by introducing semantic aspects of opcode analysis that go beyond frequency statistics in [3]. In particular, the studies since the past decade have employed Term Frequency (TF) and Term Frequency with Inverse Document Frequency (TF-IDF) as model features. The classification algorithms used in the research included SVM, Logistic Regression, Decision Trees and Random Forests, Artificial Neural Networks, Naive Bayes, and their boosted versions. The proposed method achieved 96% accuracy with a machine learning classifier, Random Forest. [13] is another study, where the features were extracted from n-gram opcode sequences and used in five machine learning classification algorithms. The highest accuracy rate in this study was 91.43%. Those studies revealed high accuracy values. For example, the best accuracy in [11] was 97% with the RF algorithm. In [12],

Euh *et al.* have focused on static feature extraction from malicious files. The obtained set of features, including opcode n-gram, API calls, window entropy map, were evaluated on several tree-based ensemble models such as XGBoost, AdaBoost, Random Forest. The highest average of those feature sets was 92.5 with XGBoost. Nevertheless, the feature selection process requires preliminary steps for input data modeling, resulting in a loss of robustness and limited scope for handling obfuscation and novel malware variants. The LSTM approach and similar approaches have the advantage of learning patterns in data by adapting the changes, making them robust and easy to maintain. In the present study, we employed the DL approach as a complementary approach to the previous works that have been conducted by employing classical ML algorithms.

Another prominent feature in malware detection is API calls in Portable Executables (PEs). API calls can serve as a clue that may facilitate the investigation of the behavior of the PEs. The API calls are usually handled in two forms: sequence (string) and graph. Several methods have been used to classify the API sequences and graphs for the past decade, including string similarity, graph similarity, and machine learning classifiers. For example, in [14] the focus was on API call graphs. Since the graph matching causes problems while graphs are growing, the call graphs were converted into a new graph type called *code graph*, and then the similarity between code graphs was measured using a predefined method. The proposed method achieved a 91% detection ratio, a relatively low accuracy rate considering the amount of preprocessing, including collecting APIs, generating call graphs, and converting them into code graphs. A similar approach was proposed for API call sequences [15]. DNA sequence alignment algorithms were adopted to explore API call sequence patterns, and these patterns were used to detect malware, even their new unknown variants. The accuracy rate achieved in this study was %99.9. However, this proposed method suffers resources and time required by the DNA sequence alignment algorithms. In [16], Cheng *et al.* aimed to detect new malware variants by clustering malware families. They used API call dependency graphs of malware samples from the same family to create the family dependency graph of each family in the dataset (including six different families). The accuracy rates of the proposed method varied between 88% and 98%, and the average accuracy rate was 92%.

Numerous studies have proposed the application of DL methods. For example, Pascanu *et al.* implemented a two-step approach consisting of feature extraction and classification in [4]. They first employed Echo State Networks (ESNs) and Recurrent Neural Networks (RNNs) to extract features from API call sequences. Next, at the classification step, there exists Logistic Regression and Multi-Layer Perceptrons with Rectifier Units. In contrast, Kolosnjaji *et al.* [17] used system call sequences to classify malicious files as malware families. They proposed a combined architecture of convolutional and recurrent LSTM layers to achieve the best results at classification. With their final model, they got an 89.4% average accuracy rate while classifying samples into families. In [18], there were three different malware classification architectures. Two of these employed language modeling built on specialized RNN architectures, LSTM, and Gated Recurrent Unit (GRU). The collected system call sequences were used to create language models with LSTM and GRU independently. The authors also proposed the third architecture based on Character-level Convolutional Neural Networks (CNN) to classify malware. The LSTM was the model with the best performance. Unlike the previous API call studies, in [19] and [20], the API calls and the input parameters used during the calls were worked together. In [19], an LSTM model was proposed to categorize files into malicious and benign categories. Zhang *et al.* [20] extracted features from API calls and their associated parameters. The API data was passed through multiple Gated-CNNs to select essential and relevant information. Then the output of Gated-CNNs was concatenated and given bidirectional LSTM to learn patterns in API calls. The highest accuracy rate in this study was 95.33%.

---

[1]In the present study, we use the term ''opcode'' to mean a single instruction that can be executed by the CPU.

Several recent studies have employed deep learning for malware detection by opcode and assembly code. In [21], Khan *et al.* focused on cancer prediction and malware detection tasks together using Convolutional Neural Networks (CNN). Like the X-Ray images for cancer prediction, "opcode pictures" were generated from opcode sequences of malicious and benign binary files. They have done each classification task on four different ResNet models, and while the best accuracy for cancer prediction was 98%, it was 88.36% for malware detection. In [22], Khan *et al.* investigated GoogleNet and five different ResNet models by using images produced from opcodes of binary files. Histogram standardization enlargement and disintegration techniques were used to upgrade images to make the differences between malicious and benign opcode images easily detectable. The accuracy rate of GoogleNet was 74.5%, and the best accuracy rate among ResNet models was 88.36%. In [23], Kumar *et al.* employed CNNs to classify malware opcode images. The accuracy rate of correctly classified binary files was 98%. However, Convolutional Neural Networks are not strong enough against small changes in images, as shown in [24] and [25]. Thus, obfuscation techniques used by malware can easily change the images generated from malware opcodes, which may cause a problem with this type of detection method. In the present study, we preferred to apply text classification approaches that are more resistant to data changes.

Furthermore, in [26], Jahromi *et al.* proposed a stacked LSTM method with pre-training of the neural network to avoid problems caused by random initialization. The final proposed LSTM model, consisting of four layers, evaluated on six different malware datasets, including static and dynamic features of Windows, Android, and IoT malware. One of the datasets was statically collected opcodes of Windows binaries, and the proposed method achieved an 88.51% accuracy rate on this dataset. As a difference from this study, we investigated the success of dynamically generated data and, instead of working on opcode, we used whole assembly output to train our neural network. While the data from the dynamic analysis in our study increases the detection rate, using assembly output in ISM without preprocessing and in BSM with small preprocessing to change the format of data decreases the required time to classify a binary file. In [7], Tang *et al.* proposed a 2-layer LSTM architecture and used the whole assembly code without just picking opcodes. They changed the representation of the binary data stream by transforming every 8 bits into an unsigned integer. Then, they trained and tested the neural network on the integer sequences. The model achieved an 89.6% accuracy rate. Tang *et al.*'s study is the closest to our study since they focused on the whole assembly code and used LSTM to create models. The difference in our study is that we focused on dynamic analysis data instead of statically collected data. We did not do any preprocessing after collecting the data in ISM and did a small amount of preprocessing to put the data in a different format. Also, we achieved a similar accuracy rate with our ISM model and got better results with the BSM model.

A further review of the literature reveals that LSTM models may have a better performance than standard RNN models in NLP applications [7] and [8]. An LSTM model of English and French languages achieved 8% improvement in perplexity over standard RNN language models [7]. In [8], Sundermeyer *et al.* compared count-based models to feedforward, recurrent, and LSTM neural networks on two large-vocabulary recognition tasks. As a result of the comparison, RNNs outperformed Feedforward Neural Networks, and standard RNNs were outperformed by LSTM with 14% reduction in English development data. Also, [28] employed LSTM architecture on "Google's One Billion Word Benchmark" dataset. This study's best improvement is a reduction in perplexity from 51.3 to 30 with combined CNN and LSTM architectures. Furthermore, the best LSTM architecture exhibited the best performance on rare words than other models implemented in the study. In another language modeling exploration [29], a language modeling was established using Czech spontaneous phone calls and the Wall Street Journal corpus to compare results with well-known data. The following section presents the methodology of the present study.

## III. METHODOLOGY

In this section, we first describe our approach for malware detection. Then, we describe how we created datasets. Finally, we introduce the architecture for dynamic malware analysis.

### A. APPROACH
Malware detection methods are usually divided into two classes: static approaches and dynamic approaches. While static approaches use PE files without executing them, dynamic approaches focus on data dynamically generated by malware during execution. In the early times of malware detection, dynamic approaches investigated dynamically generated data to extract signatures in different forms, such as string and graph. However, techniques such as obfuscation made traditional methods ineffective. Besides, the growing populations of malicious software required automated systems to detect malware. The focus shifted to Machine Learning (ML) classifiers in the studies that employed dynamic approaches. Even if ML classifiers made it possible to create automated systems to detect malware, those methods were still limited. Dynamic approaches with ML classifiers suffered from feature extraction, which caused the proposed dynamic approach methods to be less effective against new malware variants and obfuscated versions of known ones. In the present study, we apply dynamic approaches to neural networks to establish an automated system. Our proposed method is different from traditional dynamic approaches since it does not require signature feature extraction, making it a better candidate to detect new malware variants and obfuscated ones.

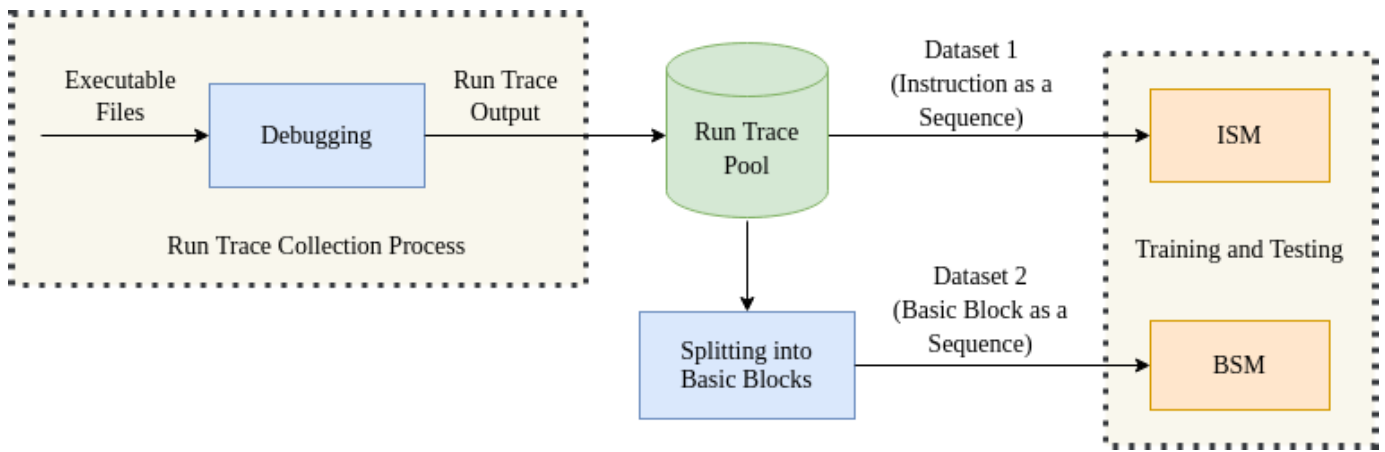

**FIGURE 1.** The data processing pipeline.

We focus on assembly instructions[2] processed during execution for malware detection. Our proposed methodology's first step is to execute each benign/malicious file in a debugger to obtain run trace outputs. Next, the outputs are saved in plain text files such that each line includes one assembly instruction, namely ''Instruction as a Sequence Model'' (viz. ISM). Then, the first dataset is processed and a second dataset with one basic block[3] per line is obtained. We call this model ''Basic Block as a Sequence Model'' (viz. BSM). Finally, we feed our LSTM (Long Short-Term Memory) language modeling architecture [6] with our datasets. The overall processing pipeline is shown in Figure 1.

The rationale behind the present methodology is to apply deep learning methodologies, which have been used for NLP (Natural Language Processing) modeling, to classify run traces of malicious and being executables. There are similarities between a natural language and the assembly language that allow the application of NLP techniques to model run traces of the assembly language. More specifically, specific grammatical rules of natural language exhibit similarities to intricate patterns exhibited by assembly instructions. The standard meaningful unit in many natural languages is the concept of word, which has a functional role similar to the opcodes and operands[4] in the assembly language. Instructions of the assembly language perform certain, modular operations in a similar way that words convey meaningful units through modular phrases and sentences in a natural language. Moreover, paragraphs of a natural language may be conceived as sharing certain characteristics with the assembly language's basic blocks. Our study investigated both the assembly instructions (in the ISM model) and the basic blocks (in the BSM model) in two separate datasets. In the following section, we present the datasets.

### B. THE DATASETS

We designed and developed two datasets for the study: the sequences of instructions (for the ISM model) and the basic blocks (for the BSM model). We obtained native x86 PE (Portable Executable) files from Windows Operating Systems (Microsoft Windows 8.1 Pro (OS Build 9600), Microsoft Windows 10 Pro 19.09 (OS Build 18363.418), and Commando VM v-2.0 [30]). Malicious executables were downloaded from the VirusShare website [31]. Since we aim to detect malware, we randomly chose the malicious samples, including various malware types, such as virus, worm, and trojan.

The datasets consist of run trace outputs, which are the sequences of the assembly instructions resulting from executing the Portable Executable (PE) files. Table 1 reveals that a total of 290 PEs were used to design a language model and conduct experiments. We processed 141 malicious PEs that consisted of 188 million instructions and 43 million basic blocks. As for the benign files, 149 PEs consisted of 151 million instructions and 14 million basic blocks.

---

[2]In the present study, we use the term ''assembly instruction'' to mean expressions consisting of opcodes and operands.

[3]In the present study, we use the term ''basic block'' to mean a piece of straight-line code which has no branch in or out except entry and exit point of a block.

[4]In the present study, we use the term ''operand'' to mean the arguments of an instruction. The data for a source operand can be found in the following locations: a register, a memory, an immediate value, and an I/O port. When an instruction returns data to a destination operand, it can be returned to: a register, a memory location, and an I/O port.

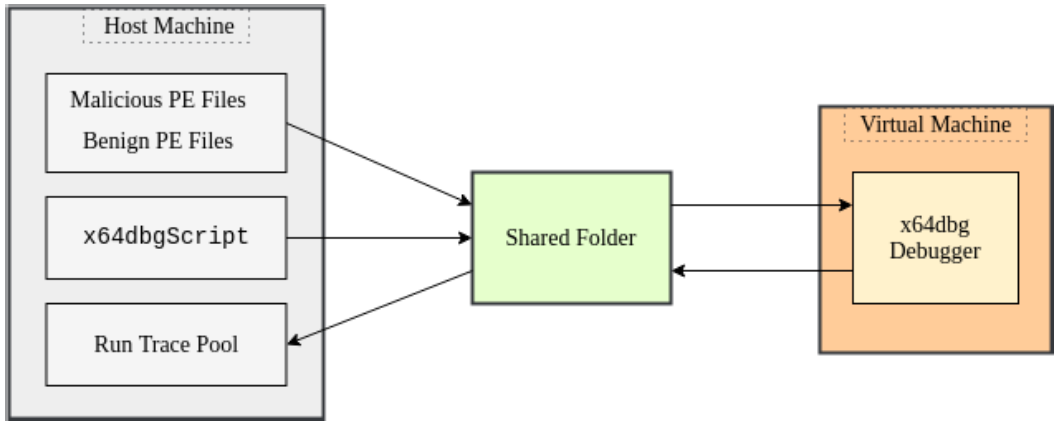

**FIGURE 2.** The run trace collection process (`MainScript`).

**TABLE 1.** Characteristics of datasets (M is the abbreviation for million).

|  | Malicious | Benign | Total |
|---|---|---|---|
| Number of Instructions in Dataset 1 for ISM | 188 M | 151 M | 339 M |
| Number of Basic Blocks in Dataset 2 for BSM | 43 M | 14 M | 57 M |
| Number of Files | 141 | 149 | 290 |

Figure 2 depicts the run trace collection process employed in the present study. We collect run trace outputs of each binary file by executing them on a debugger in a 32-bit Windows XP Professional Service Pack 3 virtual machine for creating the dataset. For this, a Windows XP virtual machine is initially prepared, and a snapshot is taken with VirtualBox (version 5.2.34) [32]. We wrote a bash script called `MainScript`[5] that runs on the host system. `MainScript` handles all PE files one by one from the input folder and repeats the following steps for each file, as shown in Figure 2. The x64dbg debugger [33] is seen as ready to use when the virtual machine is restored from the initial snapshot and started. We keep the benign and malicious PE files on the Linux host machine (Ubuntu 18.04.4 LTS). An `x64dbgScript` is generated for the corresponding executable file. The executable file and corresponding `x64dbgScript` are moved into the shared folder, which serves as a bridge between the host machine and the virtual machine. The virtual machine is restored from the snapshot and started. The `MainScript` goes on standby on the host computer until the x64dbg debugger window is closed on the virtual machine. At this point, we manually load the `x64dbgScript` to the debugger from the shared folder. After the script is run on the debugger, we wait until the executed PE file halts, or the maximum executed instruction limit is reached, which is set as 10 million. While the `x64dbgScript` is running, there might be exceptional situations such as invalid PE files, so we observe the process and intervene if necessary. Before moving to the next file, the generated run trace output saved in a text file is moved from the shared folder into the host machine's run trace pool. The `MainScript` continues until all files in the input folder are processed.

[5] https://github.com/malwareanalysislab/malware-detection-runtrace.git

During the execution of the PEs, some instructions come from the system libraries and others from the user code space. Since the system libraries' instructions are common for both malicious and benign files, we process the code section instructions from each PE file.

We analyzed two different formats of the same run trace output (one for ISM and the other for BSM). For ISM, we worked on the plain version of the run trace output, i.e., per-line instruction. Preprocessing was not required for this format, as the run trace outputs obtained from the debugger were used directly in the modeling phase. Sample lines from the dataset of the ISM are shown in Figure 3.

```
1  mov edi, eax
2  add esp, 0xC
3  test edi, edi
4  jne 0x00428817
5  mov eax, edi
```

**FIGURE 3.** Sample lines from the ISM dataset.

Next, we converted the first version (for ISM) to the basic block per sequence format to get our BSM dataset. Sample basic blocks from the second dataset are shown in Figure 4.

We wrote a Python script to parse the run trace output into basic blocks. The assembly instructions of the ISM dataset were input into the script. The script splits the run trace output text file from the basic block endpoints (i.e., branching points) by scanning it from beginning to end. This process was repeated for each run trace output text file of PEs. The script used three categories of opcodes that terminate a basic

```
1  mov esi, dword ptr ds:[0x00401180] mov edx, eax lea ecx, ss:[ebp-0x30] call esi
2  mov ebx, dword ptr ds:[0x00401168] lea ecx, ss:[ebp-0x34] push eax push ecx call ebx
3  mov eax, dword ptr ds:[0x0040A550] or eax, eax je 0x00402387 jmp eax
4  mov dword ptr ss:[ebp-0x48], eax call dword ptr ds:[0x00401054]
5  lea edx, ss:[ebp-0x34] lea eax, ss:[ebp-0x30] push edx push eax push 0x2 call dword ptr ds:[0x00401148]
```

**FIGURE 4.** Sample lines from the BSM dataset.

block: unconditional and conditional branches (e.g., "jmp, jz, jnz, jb, jl, jle, jnb, jbe, jge, ja, jns, js, je"), return instructions (e.g., "ret"), function calls (e.g., "call").

In summary, we processed the two different datasets obtained from the run trace outputs for two different models: ISM and BSM. The proposed models are the subject of the next section.

## C. THE MODELS

This section introduces two malware detection models with Long Short-Term Memory (LSTM), a specialized type of Recurrent Neural Networks (RNN) architecture. RNN architecture differs from its predecessors due to its ability to remember previous situations. However, since it has a short memory, performance problems are encountered when using RNN while processing long sequences. There are also vanishing and exploding gradient issues in the standard RNN architecture. LSTM, a special kind of RNN architecture, solves the gradient problems and improves standard RNN by modifying the cell structure. Several studies (e.g [7] and [8]) showed that LSTM is successful in extracting semantics from sequential data. Also, in malware detection, opcode sequences and API sequences are modeled by LSTM and achieved good results (e.g., [17] and [34]). In our study, a text classification neural network, such as an RNN on sequential data, is the best choice since we approach the malware detection problem from an NLP perspective. As previously noted, among RNN and its specialized types, LSTM shows better performance by minimizing the vanishing and exploding gradient issues. Thus, we applied the LSTM model to the datasets presented above. In particular, we employed LSTM for modeling the assembly codes of the PE files that already exhibit sequential structures that involve semantic relations.

We propose a six-layer architecture for the two models (ISM and BSM) in a sequential structure, as shown in Figure 5. The first layer of the architecture is the embedding layer. At this layer, the word embedding vectors are created. The second layer is the bidirectional LSTM layer. After the LSTM layer, Global Max Pool is used in the pooling layer to reduce the size of the vectors. The dropout layer is added after the pooling layer to activate selected nodes in the network to increase learning efficiency. The first dense layer reduces 128 dimensions to 64 dimensions. The second dense layer reduces 64 dimensions to 2 dimensions. At this last layer, a sigmoid is used as an activation function since the problem we work on requires to perform binary classification.

We implemented Algorithm 1 on Python 3.6.5 by using TensorFlow [35] and Keras [36] open-source libraries.

**Algorithm 1:** Algorithm for Modeling

**input** : $runTracePool = \{f_1, f_2, \ldots, f_N\}$ where $N = 290$
**output:** *accuracyRate*, *loss*

**for** $f \in runTracePool$ **do**
  **if** $f$ *is malware* **then**
    *currLabel* ← malw;
  **else**
    *currLabel* ← bngn;
  **for** $l \in f$ **do**
    *sequences.value* ← Append($l$);
    /* Append each line $l$ of file $f$ to the list *sequences*. */
    *sequences.label* ← Append(*currLabel*);
    /* Append label of each line $l$ to the list *sequences*. */

**for** $s \in sequences$ **do**
  *ts* ← Tokenize($s$);
  *tokenizedSequences* ← Append(*ts*);

**for** $ts \in tokenizedSequences$ **do**
  *ps* ← Padding(*ts*);
  /* *ts* is padded with "0" up to the maximum sequence length (8 for ISM or 30 for BSM). */
  *paddedSequences* ← Append(*ps*);

(*trainSet*, *testSet*) ← Split(*paddedSequences*);
/* The dataset is splitted into two as 80% training set and 20% test set. */
(*trainSet*, *validationSet*) ← Split(*trainSet*);
/* 25% of the train set is reserved for the validation set. */
Train(*trainSet*);
(*accuracyRate*, *loss*) ← Test(*testSet*);

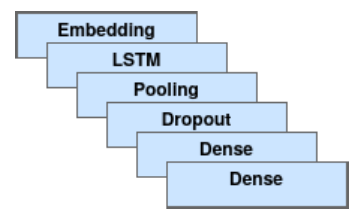

**FIGURE 5.** The layers of our proposed architecture.

So far, we have described our algorithm for the LSTM model and presented the proposed architecture.

The parameters for model training are presented next. To find the best parameter combination, we conducted experiments on the datasets. In particular, we manipulated values for a specific set of parameters, including *maximum sequence length*, *dropout rate*, *optimizer*, and the *number of LSTM nodes*. In our research, the maximum sequence length is the only parameter that takes different values for both ISM and BSM. The remaining parameters are kept the same between the models.

For the ISM modeling, we developed four models with different maximum sequence length parameter values. The optimal value was 8 for this parameter. For the BSM, we developed five models with different maximum sequence length parameter values. The optimal value was 30. We developed three models with different values for the dropout rate. The optimal value was 0.2 for both ISM and BSM. We tested five different optimizers on the models. The *Adam* optimizer [37] was the best one. We developed four models to find the optimal number of LSTM output nodes. The best value was 64. Finally, we investigated how many epochs our models (ISM and BSM) needed to learn using the best parameter values. Twenty was the highest number of epochs we used. The neural network learned from the data mostly in the first three epochs. After three epochs, there was neither a noticeable decrease in loss nor an increase in accuracy rate. We decided to train the neural network for three epochs as the lesser number of epochs reduces the risk of overfitting.

## IV. RESULTS

In this section, we present the results of the two models, namely the ISM and the BSM.

### A. THE ISM (Instruction AS A SEQUENCE Model)

We conducted a total of 16 experiments for the first model, ISM by manipulating 4 values for the *sequence length*, 3 values for *dropout rate*, 5 values for *optimizer* and 4 values for the *number of LSTM nodes*. The resulting number of correctly and incorrectly classified samples are shown in a confusion matrix (Table 2). The number of true negatives *TN* in the confusion matrix refers to correctly recognized instructions as benign instructions. In contrast, the number of true positives *TP* refers to correctly recognized instructions as malicious instructions. The number of false positives *FP* shows benign instructions recognized as malicious, whereas the number of false negatives *FN* shows malicious instructions recognized as benign.

The true positive rate *TPR* is calculated by (1) as 92.19% and the false positive rate *FPR* is calculated by (2) as 18.34%. Accuracy rate *ACC* is calculated by (3) as 87.51%.

$$TPR = \frac{TP}{(TP + FN)} \quad (1)$$

where *TP* is the number of True Positive cases and *FN* is the number of False Negative cases.

$$FPR = \frac{FP}{(FP + TN)} \quad (2)$$

where *FP* is the number of False Positive cases and *TN* is the number of True Negative cases.

$$ACC = \frac{(TP + TN)}{(TP + TN + FP + FN)} \quad (3)$$

### B. THE BSM (Basic BLOCK AS A SEQUENCE Model)

For the second model, namely BSM, we conducted 17 experiments by manipulating 5 values for the *sequence length*, 3 values for the *dropout rate*, 5 values for the *optimizer* and 4 values for the *number of LSTM nodes*.

After training the BSM, we evaluated it on the test set, which consisted of approximately 11 million basic blocks. The resulting number of correctly and incorrectly classified samples are shown in Table 3 in the confusion matrix.

**TABLE 2.** Confusion matrix of test set from the evaluation process of ISM where *TN* is the number of true negatives, *FN* is the number of false negatives, *FP* is the number of false positives, and *TP* is the number of true positives.

| | Predicted Class | |
|---|---|---|
| Actual Class | Malware | Benign |
| Malware | TP 34,810,727 | FN 2,951,034 |
| Benign | FP 5,544,678 | TN 24,691,072 |

**TABLE 3.** Confusion matrix of test set from evaluation process of BSM where *TN* is the number of true negatives, *FN* is the number of false negatives, *FP* is the number of false positives, and *TP* is the number of true positives.

| | Predicted Class | |
|---|---|---|
| Actual Class | Malware | Benign |
| Malware | TP 8,705,965 | FN 13,816 |
| Benign | FP 70,625 | TN 2,628,383 |

The true positive rate *TPR*, calculated using (1), was 99.84% and the false positive rate *FPR*, calculated using (2), was 2.62%. Finally, the correctly classified percentage of samples, accuracy rate *ACC*, calculated using (3), was 99.26%.

### C. COMPARISON OF THE MODELS

Table 4 summarizes the findings obtained by the two models.

The only factor that led to the differences between the two proposed models was the data processing format.

**TABLE 4.** Evaluation of our proposed models.

|  | **TPR**(%) | **FPR**(%) | **ACC**(%) |
|---|---|---|---|
| ISM | 92.19 | 18.34 | 87.51 |
| BSM | 99.84 | 2.62 | 99.26 |

The assembly instructions are the fundamental part of an executable's source code. It includes meaningful information and patterns and allows the ISM to achieve an 87.51% accuracy rate. However, the basic blocks in assembly code consist of more than one instruction, which is functionally related. Besides the words in instructions, there are also relations between different instructions in a basic block. Thus, the basic blocks with their longer and more complex structures include more meaningful information and more patterns than instructions, resulting in 99.26% accuracy rate in the BSM. In summary, the experiment results on the two models suggest that the basic block as a sequence model (BSM) representation exhibits a better structure for LSTM modeling compared to the instruction as a sequence model (ISM) representation.

## V. DISCUSSION

In the present study, the accuracy rates have been reported for assembly code sequence classification. That is because we aimed to separate assembly sequences as benign and malicious. In contrast, in the literature, the accuracy rates have been usually reported for the classification of files as malware or benign. This situation leads to a difference in the meaning of accuracy in the present study and the reported accuracy for file-based malware classification. Keeping this difference in mind, we compare the accuracy values below to evaluate the proposed methods' performance.

The studies that focused on opcode sequences and assembly codes can be divided into three main categories: Classical Machine Learning Classifiers, Convolutional Neural Network Classifiers, and Recurrent Neural Network Classifiers. Also, the majority of those studies focus on opcode sequences while a few others investigate assembly code as a whole.

In [3], [11] and [14], a set of machine learning classifiers were evaluated, as shown in Table 5. Random Forest methods returned the highest accuracy rates. The accuracy rate obtained in [11] is only slightly different than the proposed BSM method in the present study. However, feature extraction is a challenge in classical machine learning models that use opcodes. These models are also not effective in handling the obfuscation methods that change the opcodes' statistical information. A neural network does this feature extraction operation effectively. Therefore, neural network models may be conceived as a better solution than the classical ML models in similar performance cases.

The proposed BSM method revealed higher accuracy rates than the CNN - ResNet 152 [21] and CNN - GoogleNet [22] by focusing on assembly code sequences, which include more semantic relationships, instead of opcode sequences. The BSM model also achieved a slightly better accuracy rate than the CNN method proposed in [23], which also focused on the whole assembly code. Also, text classification neural networks do not require converting opcode or assembly code into an image as CNN, which reduces the required preprocessing time.

In [27] and [26], two LSTM architectures were reported, which used two and four hidden LSTM layers, respectively. The proposed ISM method achieved similar accuracy rates with those two models. Our BSM method improved the accuracy rate by approximately %10 by achieving %99.26. Our LSTM architecture included a single hidden LSTM layer, which made the architecture less complicated.

Finally, we developed a novel method to detect malicious code by extending previously proposed methods in several aspects. Our method is capable of detecting malware using obfuscation techniques by employing dynamic analysis since run trace outputs are dynamically generated. The ability of neural networks to adapt to data changes makes our method robust against malware obfuscation methods. In addition, working on assembly code without preprocessing in ISM and with small preprocessing in BSM keeps the time minimum spent for preprocessing. Also, using a single LSTM layer reduces the need for resources required for training the model.

## VI. LIMITATIONS AND FUTURE WORK

A major limitation of the study is that some of the steps in the pipeline require human intervention. The malware and benign files' coverage is limited to a few hundred files, despite the rich dataset obtained with the run trace collection. Future research should address improvements of the data processing pipeline; in particular, an API is needed for the x64dbg debugger. This will allow collecting data automatically for a given x86 Windows executable file and improving the run trace collection process by modifying the executable files that run on multiple CPUs in parallel. Future research should also address moving our current detection process from the code level to the file level and applying our proposed method for classifying different types of malware, such as worms, trojan horses at the OS level both for desktop and mobile operating systems.

## VII. CONCLUSION

In this paper, we proposed a novel approach to classify malicious and benign executables. We worked on assembly language. Unlike other studies using *opcodes*, we implemented our approach on dynamic analysis of run traces instead of static analysis. With the deep learning architecture LSTM, we modeled malicious and benign software run traces like natural languages.

**TABLE 5.** Evaluation of our proposed methods.

| Method | Data Format | ACC(%) |
|---|---|---|
| Random Forest [3] | Opcode | 96.00 |
| Random Forest [11] | Opcode | 97.00 |
| Random Forest [13] | Opcode | 91.43 |
| CNN - ResNet 152 [21] | Opcode | 88.36 |
| CNN - GoogleNet [22] | Opcode | 74.5 |
| CNN [23] | Instruction | 98.00 |
| 4-layer LSTM [26] | Opcode | 88.51 |
| 2-layer LSTM [27] | Opcode | 89.6 |
| ISM | Instruction | 87.51 |
| **BSM** | **Basic Block** | **99.26** |

The neural networks were trained on two datasets with different representations of the same run trace outputs. In one of the datasets, the sequences were structured as instructions. In the other dataset, the sequences were structured as basic blocks. The ISM model was trained on the dataset that used an instruction per sequence. Then we designed the BSM model to achieve better accuracy. Since the basic blocks in the assembly code consisted of functionally related components, we processed run trace outputs by splitting them into basic blocks for the BSM.

We selected optimum parameter values for our neural network architectures based on experimental results. The resulting accuracy rate (87.51%) with the ISM shows that it is possible to classify malicious and benign assembly codes by LSTM. When our improved model BSM was used, 99.26% accuracy and 2.62% false positive rate were achieved, better than ISM and most of the previously reported models. Our proposed framework for the dynamic analysis of run trace data also makes the approach resistant to polymorphic and metamorphic malware.

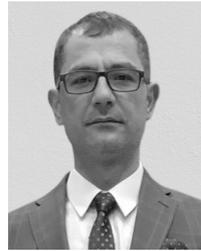

**DENIZ DEMIRCI** received the B.S. degree in system engineering from the Turkish Military Academy, Ankara, Turkey, as lieutenant, in 2007. He is currently pursuing the degree with the Cyber Security Graduate Program, Informatics Institute, Middle East Technical University (METU), Turkey.

He worked as a Platoon Leader and a Team Commander, from 2007 to 2011. From 2011 to 2016, he worked as a Software Developer. Since 2016, he has been responsible for performing and managing penetration tests, malware analysis, security incident detection and response as a Technical Lead with the Cyber Security Center of Turkish, Gendarmerie General Command. His research interests include software security, image processing, natural language processing, malware analysis, reverse engineering, and machine learning.

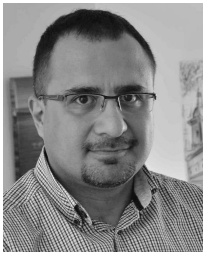

**CENGIZ ACARTURK** (Member, IEEE) received the B.Sc. degree in mechanical engineering and the M.Sc. degree in cognitive sciences from the Informatics Institute, Middle East Technical University (METU), Turkey, in 1998 and 2005, respectively, and the Ph.D. degree in computer science from the Center for Intelligent Systems and Robotics (ISR), Department of Informatics, Knowledge and Language Processing Institute (WSV), University of Hamburg, Germany, in 2010.

He was a Postdoctoral Fellow with the Microsoft Language Development Center (MLDC), Microsoft Portugal, Lisbon, from 2011 to 2012. Since 2012, he has been a Faculty Member with the Cognitive Science and Cybersecurity Graduate Programs, METU Informatics Institute. His current research interests include interaction between cybersecurity and cybercognition, human factors in cybersecurity, and natural language processing (NLP). He has been a member of ACM and the Cognitive Science Society, as well as local ICT NGOs, Turkey.

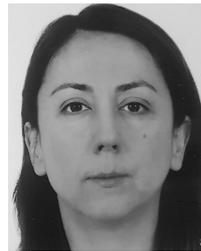

**NAZENIN SAHIN** received the B.S. degree from the Department of Mathematics, Hacettepe University, Ankara, Turkey, in 2004. She is currently pursuing the degree with the Cyber Security Graduate Program, Informatics Institute, Middle East Technical University (METU), Ankara.

From 2007 to 2012, she worked as a Software Developer, then as a Researcher with the Department of Research and Development, Turkish General Commandership of Gendarmerie, Ankara, from 2012 to 2016. Since 2017, she has been a Cybersecurity Specialist with the Turkish General Commandership of Gendarmerie. Her research interests include variety of topics in informatics, including security software development, malware analysis, reverse engineering, and machine learning in security.

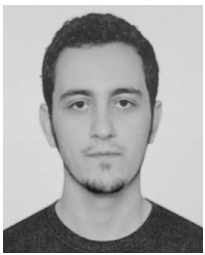

**MELIH SIRLANCI** received the B.S. degree in computer engineering from Ege University, in 2017. He is currently pursuing the degree with the Cyber Security Graduate Program, Informatics Institute, Middle East Technical University (METU), Ankara, Turkey.

Since 2019, he has been a Research Assistant with the Informatics Institute, Middle East Technical University (METU). His research interests include malware analysis, reverse engineering, binary code analysis, and machine learning.

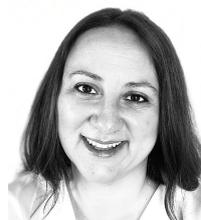

**OZGE ACAR KUCUK** received the B.Sc. degree in mathematics, and the M.B.A. and M.Sc. degrees in cryptography from Middle East Technical University (METU), Ankara, Turkey, in 2004, 2006, and 2012, respectively. She is currently pursuing the Ph.D. degree in cryptography with the Graduate Program, Institute of Applied Mathematics, Middle East Technical University (METU).

She is an Adjunct Faculty Member with the Cyber Security Graduate Program, Informatics Institute, Middle East Technical University (METU). Her research interests include cryptography, cryptanalysis, reverse engineering, malware analysis, mobile security, and vulnerability research and forensics.

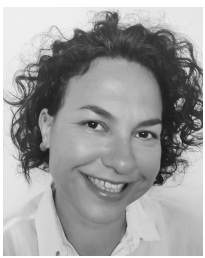

**PINAR GURKAN BALIKCIOGLU** received the B.S. degree in statistics from Hacettepe University, Ankara, Turkey, in 2004, and the Ph.D. degree in cryptography from Middle East Technical University (METU), Ankara, in 2016.

Since 2017, she has been an Adjunct Faculty Member with the Cyber Security Graduate Program, Informatics Institute, Middle East Technical University (METU). Her research interests include cryptography, reverse engineering, malware analysis, forensics, and vulnerability research and mobile security.